\documentclass{PoS}

\usepackage{enumerate}
\usepackage{amsmath,amssymb}  
\usepackage{bm}               
\usepackage{amscd}            
\usepackage{graphicx}         
\usepackage[export]{adjustbox}
\usepackage{epsfig}
\usepackage{hhline,multirow}  
\usepackage{dcolumn}          
\usepackage[dvipsnames]{xcolor}
\usepackage[normalem]{ulem}   
\usepackage{mathtools}

\usepackage{slashed}

\newcommand{\mj}{M_q}
\newcommand{\mq}{m_q}

\newcommand{\id}{{\bm 1}
}

\newcommand{\lslash}{l \hspace{-0.19cm} / \,}

\title{Transversity in inclusive DIS and novel TMD sum rules
  \thanks{Preprint JLAB-THY-18-2774}
}

\ShortTitle{Transversity in DIS and novel TMD sum rules}

\author{\speaker{Alberto Accardi}\\
  Hampton University, Hampton, VA 23668, USA \\
  and Jefferson Lab, Newport News, VA 23606, USA\\
	E-mail: \email{accardi@jlab.org}}

\author{Andrea Signori
	\thanks{ORCID: http://orcid.org/0000-0001-6640-9659} \\
	Jefferson Lab, Newport News, VA 23606, USA\\
     E-mail: \email{asignori@jlab.org}}

   \abstract{A reanalysis of collinear factorization for inclusive Deep Inelastic Scattering shows that a novel, non perturbative spin-flip term associated with the invariant mass of the produced hadrons couples, at large enough Bjorken $x_B$, to the target's transversity distribution function. The resulting new contribution to the $g_2$ structure function can potentially explain the discrepancy between recent calculations and fits of this quantity. The new term also breaks the Burkhardt-Cottingham sum rule, now featuring an interplay between the $g_2$ and $h_1$ functions that calls for a re-examination of their small-$x$ behavior. As part of the calculation leading to these results, a new set of TMD sum rules is derived by relating the single-hadron quark fragmentation correlator to the fully dressed quark propagator by means of integration over the hadronic momenta and spins. A complete set of momentum sum rules is obtained for transverse-momentum-dependent quark fragmentation functions up to next-to-leading twist.}

\FullConference{XXVI International Workshop on Deep-Inelastic Scattering
  and Related Subjects (DIS2018)\\
  16-20 April 2018\\
  Kobe, Japan}

\begin{document}

\section{DIS with a jet correlator}
\label{sec:intro}

A correct treatment of the partonic kinematics in inclusive Deep Inelastic Scattering (DIS) is important for establishing the limits of applicability of QCD factorization theorems and correctly accessing Parton Distribution Functions (PDFs) from experimental data \cite{Collins:2007ph}. This is in particular true for experiments at Jefferson Lab, but has implications also for higher energy experiments.

\begin{figure}[bth]
  \centering
  (a)\includegraphics[width=0.3\linewidth,valign=t]{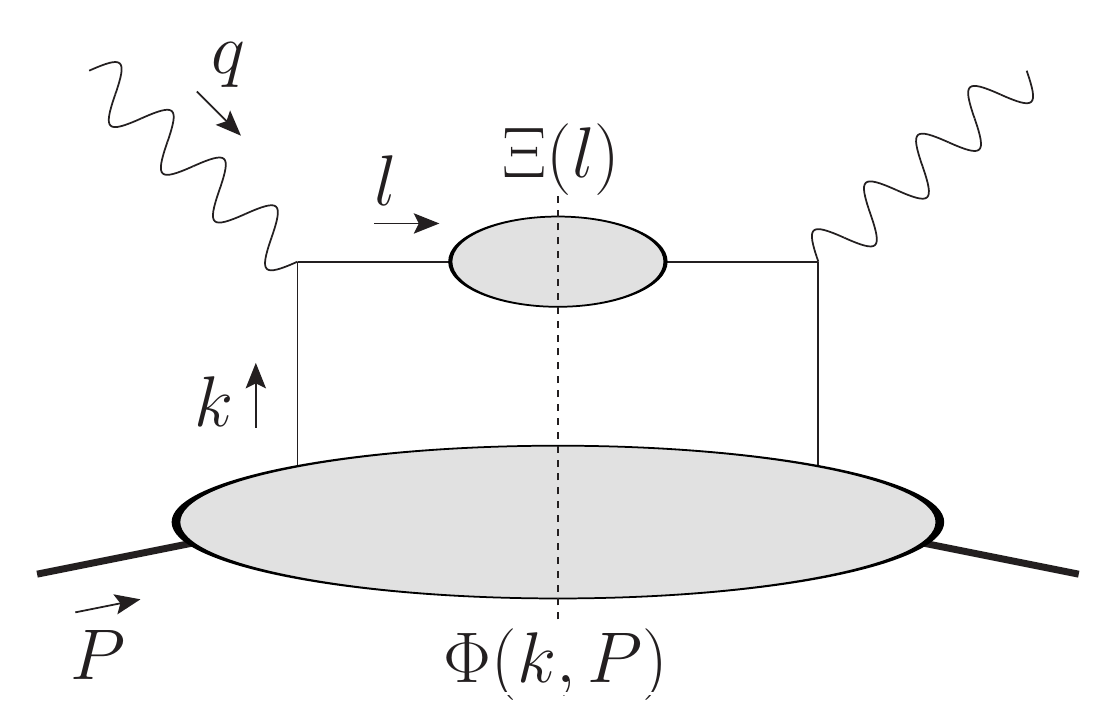}
  \hfill
  (b)\includegraphics[width=0.3\linewidth,valign=t]{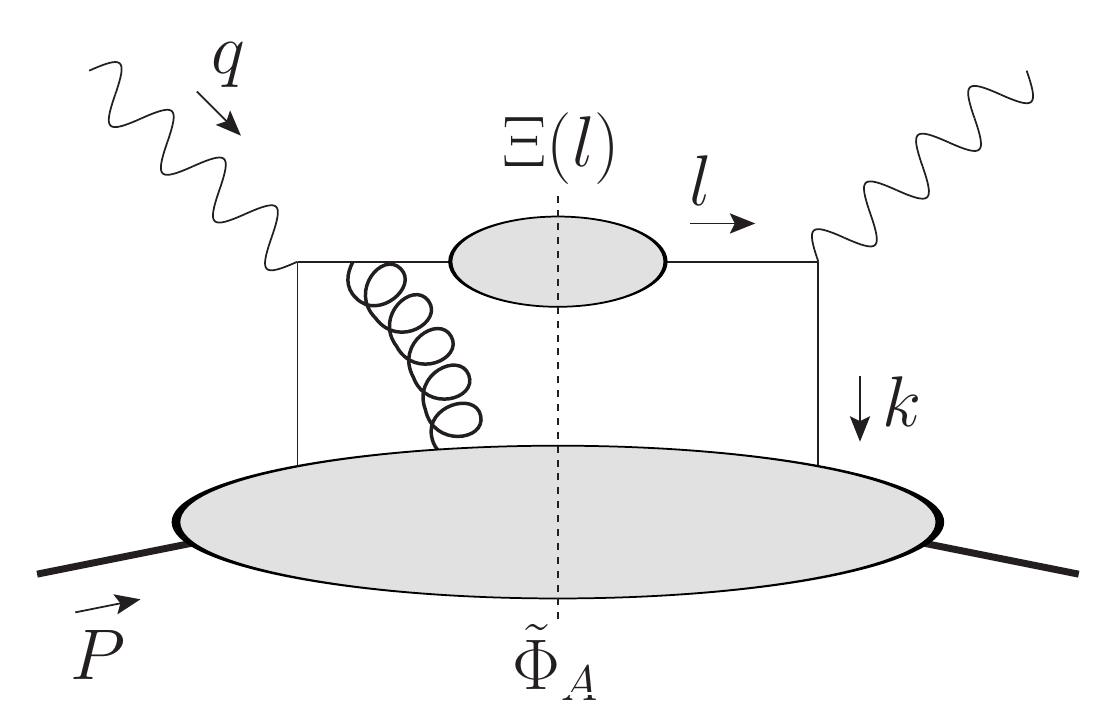}
  \hfill
  (c)\includegraphics[width=0.3\linewidth,valign=t]{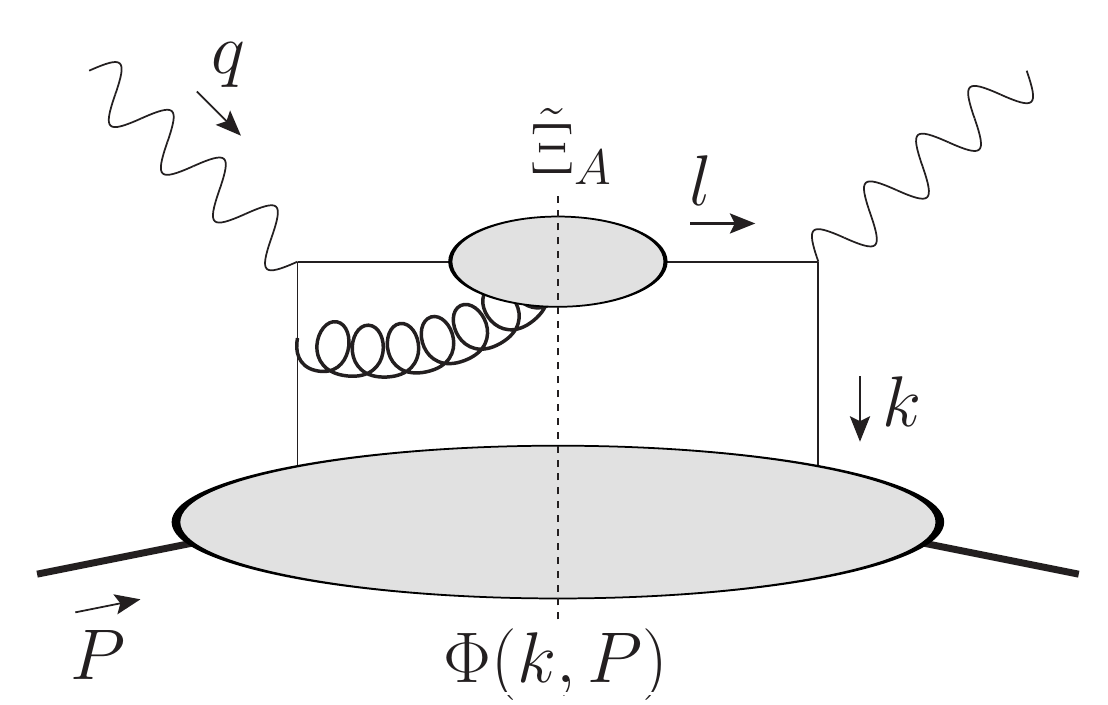}
  \caption{Jet correlators in inclusive DIS at leading order: leading twist (a), and twist-3 (b and c) diagrams.
  }
  \label{fig:handbags}
\end{figure}

Due to color confinement, scattered quarks must turn into hadrons. Thus a minimal extension of the usual handbag diagram in pQCD includes a ``jet correlator'' that describes the production of particles collinear to the scattered quark, see the top blob in Fig.~\ref{fig:handbags}a.
The appearance of a collimated jet subgraph in a fully inclusive cross section may appear at first puzzling, due to the absence of a measured scale that can physically identify the jet. Furthermore, by applying the completeness relation, it would seem that Fig.~\ref{fig:handbags}a would reduce to the usual handbag diagram with a quark line when summing over all possible final states. However, the final state invariant mass
\begin{align}
  M_X^2 = Q^2(1-x_B)/x_B + M^2
\label{eq:Mx}
\end{align}
(with $M$ the proton's mass) is kinematically limited at large $x_B$, and only a subset of final states can be physically produced. In particular, the hadrons transverse momentum is limited, and the the final state becomes more and more jet-like as $x_B$ increases.
It is the kinematics, rather than an explicit measurement, that provides the scale necessary to justify the jet subgraph. 

The leading twist diagram in Fig.~\ref{fig:handbags}a has already been analyzed in collinear factorization, and utilized to:
 \textit{(i)} perform kinematic ``jet mass corrections'' to the unpolarized $F_{1,2,3}$ structure functions, and similarly to the polarized $g_1$, using the chiral-even part of the jet correlator \cite{Accardi:2008ne}; 
 and \textit{(ii)} reveal a novel coupling of  the collinear transversity parton distribution function with the chiral-odd component of the jet correlator \cite{Accardi:2017pmi}.
The latter will be discussed in more detail in Sect.~\ref{sec:transversity}.

The quark jet correlator appearing in the top blob of Fig~\ref{fig:handbags}a is defined as \cite{Accardi:2008ne,Accardi:2017pmi}:
\begin{align}
\Xi_{ij}(l,n_+) =\int
  \frac{d^4\eta}{(2\pi)^4}\; e^{i l \cdot \eta}\,
    \langle 0|\, \mathcal W^{n_+}_{(+\infty,\eta)}
\,\psi_i(\eta)
             \bar{\psi}_j(0)\,
\mathcal W^{n_+}_{(0,+\infty)}\,   |0 \rangle \ ,
\label{e:xifull}
\end{align}
where $l$ is the quark's four-momentum, $\Psi$ the quark field
operator (with quark flavor index omitted for simplicity), $|0\rangle$ is
the nonperturbative QCD vacuum, and the Wilson line operators $\mathcal W^{n_+}$
run to infinity first along the light-cone plus direction $n_+$, then transversely to that vector~\cite{Bacchetta:2006tn}.

The correlator $\Xi$ can be parametrized in terms of jet correlation
\textit{functions} through a Lorentz covariant Dirac decomposition that utilizes the vectors $l$ and $n_+$, and is constrained by invariance under parity and time reversal:
\begin{align}
  \Xi(l,n_+) & =
  \Lambda A_1(l^2) {\bf 1} + A_2(l^2) \slashed{l}
  + \frac{\Lambda^2}{k \cdot n_+} B_1(l^2) \slashed{n}_+  \ , 
\label{e:jetexpansion-tw3}
\end{align} 
where the scale $\Lambda$ defines the power counting for the $A$ and $B$ functions.
Up to twist 3, the jet correlator is nothing else than the cut quark propagator, and the $A_1$ and $A_2$ functions can then be interpreted in terms of its spectral representation 
\begin{align} 
  \Xi(l) =  
  \int d \sigma^2 \big[ \sigma\, J_1 (\sigma^2)\,\id + J_2 (\sigma^2)\,\lslash \big] \,
  \frac{\delta(l^2 -\sigma^2)}{(2\pi)^3} \ ,
\label{e:jetspectral}
\end{align}
where $\sigma^2$ can be interpreted as the invariant mass of the current jet, and the spectral (or ``jet'') functions $J_i$ describe the jet mass distribution:
$(2\pi)^3 A_1(l^2)= \sqrt{l^2} J_1(l^2) / \Lambda$, and $(2\pi)^3 A_2(l^2)=J_2(l^2)$.
As a consequence of positivity constraints and CPT invariance, the jet functions satisfy 
\begin{align}
  J_2(\sigma^2) \geq J_1(\sigma^2) \geq 0
  \hspace*{0.5cm} \text{and} \hspace*{0.5cm}
  \int_0^\infty d\sigma^2 J_2(\sigma^2) = 1 \ .
\label{eq:jetfnsumrule1}
\end{align}
For later use, we also define the average (chiral-odd) jet mass
\begin{align}
  M_q = \int_0^\infty d\sigma^2 \, \sigma\, J_1(\sigma^2) \ ,
  \label{eq:jetfnsumrule2}
\end{align}
or, in other words, the mass acquired by the quark interacting with vacuum fluctuations. In general, $M_q>m_q$, and for light quarks we estimate $M_q=O(100 \text{\ MeV})$ \cite{Accardi:2017pmi}.

\section{Connecting single-hadron FFs and Jet Functions}
\label{sec:TMDsumrules}

In order to calculate the DIS cross section, it is useful
to integrate the SIDIS cross section, discussed up to twist 3 in Ref.~\cite{Bacchetta:2006tn}, and use a set of
momentum sum rules for single-hadron TMD Fragmentation Functions (FFs) we are going to derive here.
We take inspiration from Ref.~\cite{Collins:1981uw,Meissner:2010cc}, but connect the quark fragmentation correlator to the quark jet correlator, and extend the formalism to include pure twist-3 FFs. 
The derivation is performed in the parton frame, where the quark has transverse momentum ${\bf l}_T=0$, and the hadron has a transverse momentum ${\bf P}_{h\perp}$ relative to the quark. The results are, nonetheless, frame independent because they only involve integrated FFs.

The starting point is the correlator for the fragmentation of a quark with momentum $l$ into one hadron of momentum $P_h$ and spin $S_h$,
\begin{equation}
\label{e:1hDelta_TMDproj}
  \Delta(l,P_h,S_h) =
   \int \frac{d^4 \eta}{(2\pi)^4} e^{i l \cdot \eta } 
    \langle 0|  \mathcal W_{(\infty,\eta)}^{n_+} \psi_i(\eta) \big( a_h^\dagger(P_h S_h)a_h(P_h S_h) \big) \overline{\psi}_j(0) \mathcal W_{(0,\infty)}^{n_+} |0 \rangle \\ ,
\end{equation}
with suitably defined staple-like Wilson lines $\mathcal W$~\cite{Bacchetta:2006tn} and hadron creation and annihilation operators $a^\dagger_h$ and $a_h$. It is then possible to show that
\begin{align}
\label{e:sumrule_p1}
  \sum_{h,S_h} \int (dP_h) \,
    P_h^\mu \Delta (l,P_h,S_h) = 
  \int \frac{d^4\eta}{(2\pi)^4}e^{i l \cdot \eta}
  \langle 0|  \mathcal W_{(\infty,\eta)}^{n_+} \psi_i(\eta) \bm \hat{\bf P}^\mu \overline{\psi}_j(0) \mathcal W_{(0,\infty)}^{n_+} |0 \rangle \ ,
\end{align}
where $\bm \hat{\bf P}^\mu$ is the momentum operator~\cite{Collins:1981uw} and $(dP_h) = \frac{dP_h^- d^2 {\bf P}_{h\perp}}{2P_h^- (2\pi)^3}$ the Lorentz-invariant integration measure over the hadron momentum. Note that, as stressed in Ref.~\cite{Meissner:2010cc}, this is only possible when summing also over the hadron spin, and the ensuing sum rules are only valid for unpolarized FFs.

Next, we can make use of the eigenvalue equation in coordinate representation for the momentum operator, $\psi_i(\eta) \bm\hat{{\bf P}} = i\, \frac{\partial}{\partial \eta_\mu} \psi_i(\eta)$, and choose $\mu=-$ and $\mu=1,2$ to obtain, respectively, longitudinal and transverse sum rules. 
Integrating by parts,
and taking into account that the fields vanish at the boundary of the integration domain,
we obtain:
\begin{align}
  \sum_{h,S_h} \int (dP_h) \, P_h^\mu \Delta (l,P_h,S_h) 
    = \begin{cases}
        \ l^-\, \Xi(l) & \text{for\ } \mu=- \quad\ \ \, \text{(longitudinal)} \\
        \ 0 & \text{for\ } \mu=1,2 \quad \text{(transverse)}
      \end{cases}
\label{eq:sumrule_correlator}
\end{align}

From Eq.~\eqref{eq:sumrule_correlator}, which  is the master result of this section, we can derive sum rules for unpolarized quark TMD-FFs by taking
suitable traces with Dirac matrices. For twist-2 FFs, in the longitudinal and transverse sectors, respectively, we obtain
\begin{align}
  \quad
  && \sum_{h,S_h} \int dz\, z\, D_1(z) &= 1
  & \sum_{h,S_h} \int dz\, z\, H_1^{\perp (1)}(z) &= 0 \ ,
  &\quad
  \label{eq:sumrules-DH}
\end{align}
where $z=P_h^-/l^-$, and the superscript $(1)$ stands for the first moment in $P_{h\perp}^2$. These two sum rules were originally introduced by Collins and Soper~\cite{Collins:1981uw}, and by Sch\"afer and Teryaev~\cite{Schafer:1999kn}, respectively.
The following sum rules for twist-3 quark-quark functions are new:
\begin{align}
  \qquad
  && \sum_{h,S_h} \int dz E(z) & =\frac{M_q}{\Lambda}
    & \sum_{h,S_h} \int dz D^{\perp (1)}(z) & = 0
  \label{eq:sumrules-ED}\\
  && \sum_{h,S_h} \int dz H(z) &= 0
    &  \sum_{h,S_h} \int dz G^{\perp (1)}(z) &= 0
  &\qquad
  \label{eq:sumrules-HG}
\end{align}
with the longitudinal ones in the left column, and the transverse ones in the right column.
Finally, using equation of motion relations~\cite{Bacchetta:2006tn}, we can also obtain sum rules for pure twist 3 TMD-FFs:  
\begin{align}
  \qquad
  &&  \sum_{h,S_h} \int dz \tilde{E}(z) &= \frac{M_q-m_q}{\Lambda}
  & \sum_{h,S_h} \int dz \tilde{D}^{\perp (1)}(z) &= \frac{\langle\!\langle {\bf l}_T^2 \rangle\!\rangle}{2 \Lambda^2}
  \label{eq:sumrules-EtDt} \\
  && \sum_{h,S_h} \int dz \tilde{H}(z) &= 0
    & \sum_{h,S_h} \int dz \tilde{G}^{\perp (1)}(z) &= 0 
  &\qquad
  \label{eq:sumrules-HtGt}
\end{align}
with the longitudinal rules on the left, and the transverse ones on the right. The sum rule for $\widetilde E$ was introduced in Ref.~\cite{Accardi:2017pmi}, the others are new.
Note that $\langle\!\langle {\bf l}_T^2 \rangle\!\rangle \equiv \sum_{h,S_h} \int dz\,d^2{\bf P}_{h\perp} ({\bf P}_{h\perp}^2/z^2) D_1(z,{\bf P}_{h\perp})$ can be interpreted as the average squared transverse momentum of the quark in the hadron frame. All together, Eqs.\eqref{eq:sumrules-DH}-\eqref{eq:sumrules-HtGt} are a complete set of sum rules for unpolarized TMD-FFs up to twist-3.

\section{Transversity in inclusive DIS}
\label{sec:transversity}

It is now time to insert the jet correlator inside the DIS handbag diagram, and calculate the structure functions. In order to calculate $g_2$, we also need to account for
the quark-gluon-quark correlators in Figs.\ref{fig:handbags}(b) and (c).
The latter diagram and its Hermitian conjugate, considered for the first time in \cite{Accardi:2017pmi}, are essential to restore gauge invariance, which would otherwise be broken at twist-3 by $m_q \neq M_q$. 
Our results will be valid in a $x_B$, $Q^2$ region such that
\begin{align}
  M_\infty^2 \gg Q^2(1/x_B-1) \gg M_{\text{jet}}^2 \ ,
\label{eq:kinlimits}
\end{align}
where $M_\infty$ is the minimum invariant mass needed to utilize the completeness relation in the DIS final state, and $M_{\text{jet}}^2$ is a jet mass scale of the order of $M_q^2$ or $\int d\sigma^2 \sigma^2 J_2(\sigma^2)$. The upper bound is necessary to guarantee the relevance of the jet correlators in Fig.~\ref{fig:handbags}, and implies a smaller and smaller interval of validity in $x_B$ as $Q^2$ increases. The lower bound is only necessary to guarantee that the integrations of $J_{1,2}$ over $l^+ = l^2/(2l^-)$ extend far enough so that one can apply the spectral sum rules \eqref{eq:jetfnsumrule1}-\eqref{eq:jetfnsumrule2} and neglect jet mass corrections. Should this condition not be satisfied in actual experiments, jet mass corrections may be handled according to
Ref.~\cite{Accardi:2008ne}.

As proposed in \cite{Accardi:2017pmi}, rather than directly using the diagrams in Fig.~\ref{fig:handbags}, it is convenient to start from the semi-inclusive one, that has already been studied up to twist-3 level \cite{Bacchetta:2006tn}. Then, integrating over hadron momenta, summing over flavors and spins, and taking advantage of the longitudinal sum rules derived in Section~\ref{sec:TMDsumrules}, one obtains 
\newcommand{\xbj}{{x_B}}
\begin{gather}
  F_{T}(x_B) = \xbj \,\sum_q e_q^2\,f_1^q(\xbj),
\qquad
  F_{L}(x_B) = 0,
\qquad
  F_{LL}(x_B) =\xbj\,\sum_q e_q^2\,g_1^q(\xbj),
\label{e:FF}
\\
  F_{UT}^{\sin \phi_S}(x_B)=0,
\qquad
  F_{LT}^{\cos \phi_S}(x_B)=-\xbj \,\sum_q e_q^2\, \frac{2M}{Q}\,
                        \biggl(\xbj  g_T^q(\xbj)
                        + \frac{\mj -\mq}{M} \, h_{1}^q(\xbj) \biggr) \ ,
\label{e:FFsincos}
\end{gather}
where $f_1^q$, $g_1^q$ and $h_1^q$ are, respectively, the unpolarized, polarized, and transversity PDFs.
The vanishing of $F_{UT}^{\sin \phi_S}(x_B)= \sum_{h,S_h} \int dz F_{UT}^{h,\sin \phi_S}(x_B,z)$, known as Diehl-Sapeta sum rule~\cite{Diehl:2005pc}, is derived here for the first time at the correlator level. The second term in $F_{LT}^{\cos \phi_S}(x_B)$, that is not suppressed as an inverse power of $Q$ compared to the standard $g_T$ term, was already unveiled in Ref.~\cite{Accardi:2017pmi}.
In a perturbative calculation, one would obtain $M_q^\text{pert}=m_q$ and the new term would vanish.

\begin{figure}[tb]
\begin{center}
  \parbox[c]{0.50\linewidth}{
    \includegraphics[width=0.97\linewidth]{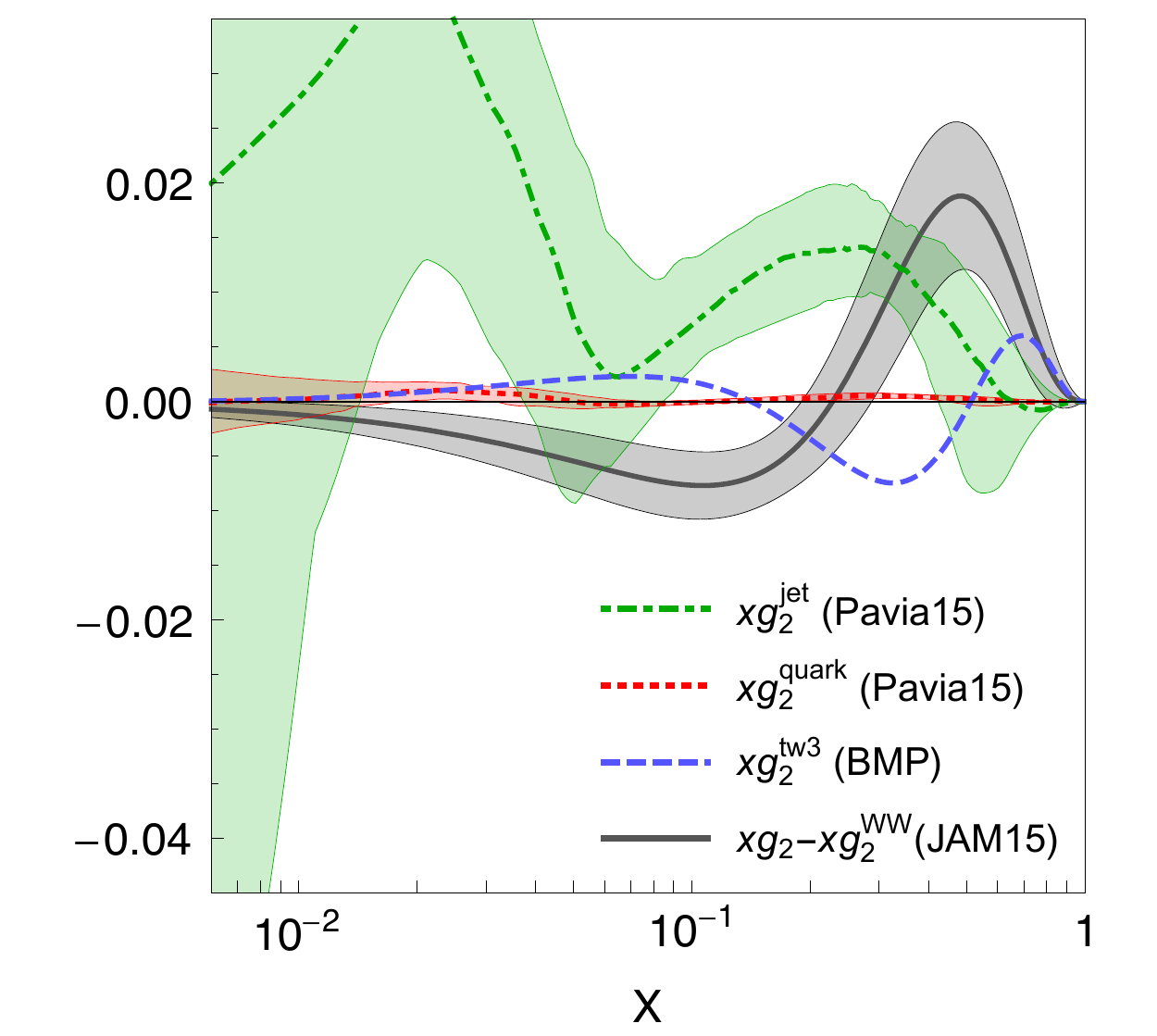} \vskip-0.7cm
  }
  \parbox[c]{0.49\linewidth}{    
    \caption{
      Different contributions to the non Wandzura-Wilczek part of the proton $g_2$ structure functions compared to the JAM15 fit of $g_2-g_2^{\text{WW}}$ (solid black)~\cite{Sato:2016tuz}. The quark and jet
contributions are shown with a dotted red and a dot-dashed green line
respectively, with uncertainty bands coming form the Pavia15 fit of the
transversity function \cite{Radici:2015mwa}. The uncertainty in the choice
$m_q=5$ MeV and $\mj=100$ MeV is not shown. The pure twist-3 piece by Braun et al. \cite{Braun:2011aw} is shown as a dashed blue line. Note that the jet contribution is plotted down to low values of $x_B$, even though one would expect it to be suppressed at $x_B\lesssim Q^2/(M_\infty^2+Q^2)$.}
  }
\label{fig:g2p}
\end{center}
\end{figure}

The new transversity-dependent  coupling also contributes to the more conventional $g_2$ structure function, as can be seen by applying the methods discussed in Ref.~\cite{Accardi:2009au}: 
\begin{align}
  g_2(\xbj) - g_2^{WW}(\xbj) =  \frac{1}{2}\,\sum_a e_a^2
\biggl(
    g_2^{tw-3}(\xbj) 
    + \frac{\mq}{M} \left(\frac{h_1^q}{x}\right)^\star\!\!\!\!(\xbj) 
    + \frac{\mj-\mq}{M} \frac{h_1^q(\xbj)}{\xbj} 
  \Biggr) \ ,
\label{eq:newg2}
\end{align}
where $f^*(x)\equiv-f(x)+\int_x^1 dy f(y)/y$, $g_2^{WW} = g_1^*$ is the Wandzura-Wilczek pure twist 2 chiral-even term, and $g_2^{tw-3}(\xbj)$ is a ``pure twist-3'' function that only depends on quark-gluon-quark matrix elements. The novelty is the last, jet mass dependent term. This is shown in Fig.~\ref{fig:g2p} to potentially have a size comparable to the other terms, although in absence of theoretical calculations or experimental determinations, we can only use a rough $M_q=O(100 \text{\ MeV})$ estimate for now \cite{Accardi:2017pmi}.

\begin{samepage}
The new term also breaks the Burkhardt-Cottingham sum rule:
\begin{align}
  \int_0^1 dx_B g_2(x_B) = \Delta_{BC} \ .
\end{align}
Trusting Eq.~\eqref{eq:newg2} down to $x_B=0$, we would obtain a divergent $\Delta_{BC} = \frac{M_q-m_q}{M}  \int_0^1dx \frac{h_1(x)}{x}$ (assuming $h_1 \propto x^{-\epsilon_g}$  with $\epsilon_g=\sqrt{\alpha_s N_c/\pi}\approx -0.56$ as for the non-singlet part of $g_1$ \cite{Kovchegov:2016zex}).
Fortunately, the kinematic limitation in Eq.~\eqref{eq:Mx} on the final state invariant mass eases off as $x_B \rightarrow 0$,
and one can use the completeness relation to recover the standard result without the new jet term.
Thus, $\Delta_{BC}$ is likely to be finite, with a $Q^2$ dependence that reflects the magnitude of the $M_\infty$ scale.

\acknowledgments

We thank A.~Metz, M.~Sievert, and I.~Stewart for helpful discussions. This work was supported by the U.S. Dept. of Energy contract DE-AC05-06OR23177, under which Jefferson Science Associates, LLC, manages and operates Jefferson Lab, and by contract DE-SC0008791.


\bibliographystyle{JHEP}
\bibliography{JSR.bib}
\end{samepage}

%

\end{document}